\documentclass[fleqn]{pos}%
\usepackage{amsmath}
\usepackage{amsfonts}
\usepackage{amssymb}
\usepackage{graphicx}%
\providecommand{\U}[1]{\protect\rule{.1in}{.1in}}

\title{%
{\protect
\vspace{-2cm}
\flushright{
\begin{minipage}{0.34\hsize}
\normalsize
CHIBA-EP-166 \\
KEK-Preprint 2007-50 \\
\end{minipage}} \\
\vspace{2cm}
}%
Toward gauge independent study of confinement in SU(3) Yang-Mills theory}
\ShortTitle{Toward gauge independent study of confinement in SU(3) YM theory%
}

\author{%
A.Shibata$^{a}$,%
S.Kato$^{b}$, %
K.-I. Kondo$^{c}$,%
T. Murakami$^{d}$,%
T. Shinohara$^{d}$%
and  S. Ito$^{e}$ \\
\llap{$^{a}$} High Energy Accelerlator Organization(KEK) \&
Sokendai,  Tsukuba, 305-0801, Japan\\
\llap{$^{b}$} Takamatsu National College of Technology, Takamatu City, 
761-8058 Japan \\
\llap{$^{c}$} Department of Physics, Graduate School of Science, 
Chiba University, 
Chiba, 263-8522 Japan \\
\llap{$^{d}$} Graduate School of Science and Technology, Chiba University, 
Chiba, 263-8522 Japan \\
\llap{$^{e}$} Nagano National College of Technology, 716 Tokuma,
 Nagano, 381-8550 Japan \\
E-mail:\email{Akihiro.Shibata@kek.jp},\email{kato@takamatsu-nct.ac.jp},
\email{kondok@faculty.chiba-u.jp},%
\email{tom@cuphd.nd.chiba-u.ac.jp}, \email{sinohara@graduate.chiba-u.jp},
\email{shoichi@ei.nagano-nct.ac.jp}
}

\abstract{

Dual superconductivity is believed to be a promising mechanism for quark confinement and has been investigated on a lattice effectively by a particular gauge called the maximal Abelian (MA) gauge. We propose a new formulation of SU(3) Yang-Mills theory on a lattice based on a non-linear change of variables where the new field variables are expected to reduce to those of the Cho-Faddeev-Niemi-Shabanov decomposition in the continuum limit. By introducing a new variable, say color field, carrying the color direction with it, this formulation enables us to restore and maintain color symmetry that was lost in the conventional MA gauge due to the naive separation of the gauge potential into diagonal and off-diagonal components. An advantage of this formulation is that we can define gauge-invariant magnetic monopoles without relying on specific gauges to investigate quark confinement from the viewpoint of dual superconductivity. In this talk, we will present the relevant lattice formulation to realize the above advantages and preliminary results of numerical simulations to demonstrate the validity of this formulation. This SU(3) formulation is an extension of the SU(2) version already proposed by us in the previous conference.
}

\FullConference{The XXV International Symposium on Lattice Field Theory\\
	July 30 - August 4 2007\\
	  Regensburg, Germany}

\begin{document}
\section{Introduction}

Quark confinement is still an unsolved and challenging problem in theoretical
particle physics. Dual-superconductivity is believed to be a promising
mechanism for quark confinement in quantum chromodynamics
(QCD).\cite{ref:DsuperCond} Indeed, the relevant data supporting the validity
of this picture have been accumulated by numerical simulations especially
since 1990 and some of the theoretical predictions\cite{ref:tHooft81}%
\cite{ref:EzawaIwasaki} have been confirmed in the Maximal Abelian (MA) gauge;
infrared Abelian dominance, magnetic monopole dominance and non-vanishing
off-diagonal gluon mass, which are the most characteristic features for dual
superconductivity. However, they are not yet confirmed in any other gauge than
the MA gauge and the MA gauge breaks color symmetry.

In this talk, we propose a new compact lattice formulation for the SU(3)
Yang-Mills (YM) theory to establish the dual superconductivity picture for
quark confinement in a gauge invariant way. This could be a lattice version of
the non-linear change of variables (NLCV) for the YM gauge field in the
continuum formulation (originally known as the Cho-Faddeev-Niemi-Shavanov
(CFNS) decomposition). There the YM gauge field $\mathbf{A}_{\mu}$ is
decomposed into two parts, $\mathbf{A}_{\mu}=\mathbf{V}_{\mu}+\mathbf{X}_{\mu
}$ in such a way that the \textquotedblleft Abelian (diagonal)" part
$\mathbf{V}_{\mu}$ is dominantly responsible for the area decay law of the
Wilson loop average, while the remaining "off-diagonal" part $\mathbf{X}_{\mu
}$ decuple in the low-energy (or long-distance) regime, thereby, leading to
the infrared Abelian dominance. For performing non-perturbative studies,
therefore, it is important to give a procedure of extracting such an
\textquotedblleft Abelian" part $\mathbf{V}_{\mu}$ and the remaining part
$\mathbf{X}_{\mu}$ from the original YM gauge field $\mathbf{A}_{\mu}$ also on
a lattice. We construct the SU(3) lattice formulation by extending the SU(2)
version proposed in our previous work
\cite{ref:Kondo,ref:NLCVsu20,ref:NLCVsu2,ref:NLCVsu2-2} (For details, see
\cite{ref:prep01}). In the SU(2) case, we have succeeded to define two compact
lattice variables $V_{x,\mu}$ and $X_{x,\mu}$which play the similar role to
the "Abelian" and "off-diagonal" parts in the continuum theory. These new
variables enable us to define a gauge invariant magnetic monopole in the
compact formulation which guarantees that the magnetic charge is
integer-valued and obeys the Dirac quantization condition. Moreover, the
infrared \textquotedblleft Abelian\textquotedblright\ dominance and magnetic
monopole dominance in the string tension were demonstrated by numerical
simulations, together with the non-vanishing mass for the "off-diagonal" part.
It is crucial to introduce a color vector field $\mathbf{n}(x)$ for maintain
the color symmetry of the original YM theory.

\section{A new compact reformulation of SU(3) YM}

\begin{figure}[ptb]
\begin{center}
\includegraphics[
bb=116 116 452 721 , height=3in, angle=-90 , origin=br
]{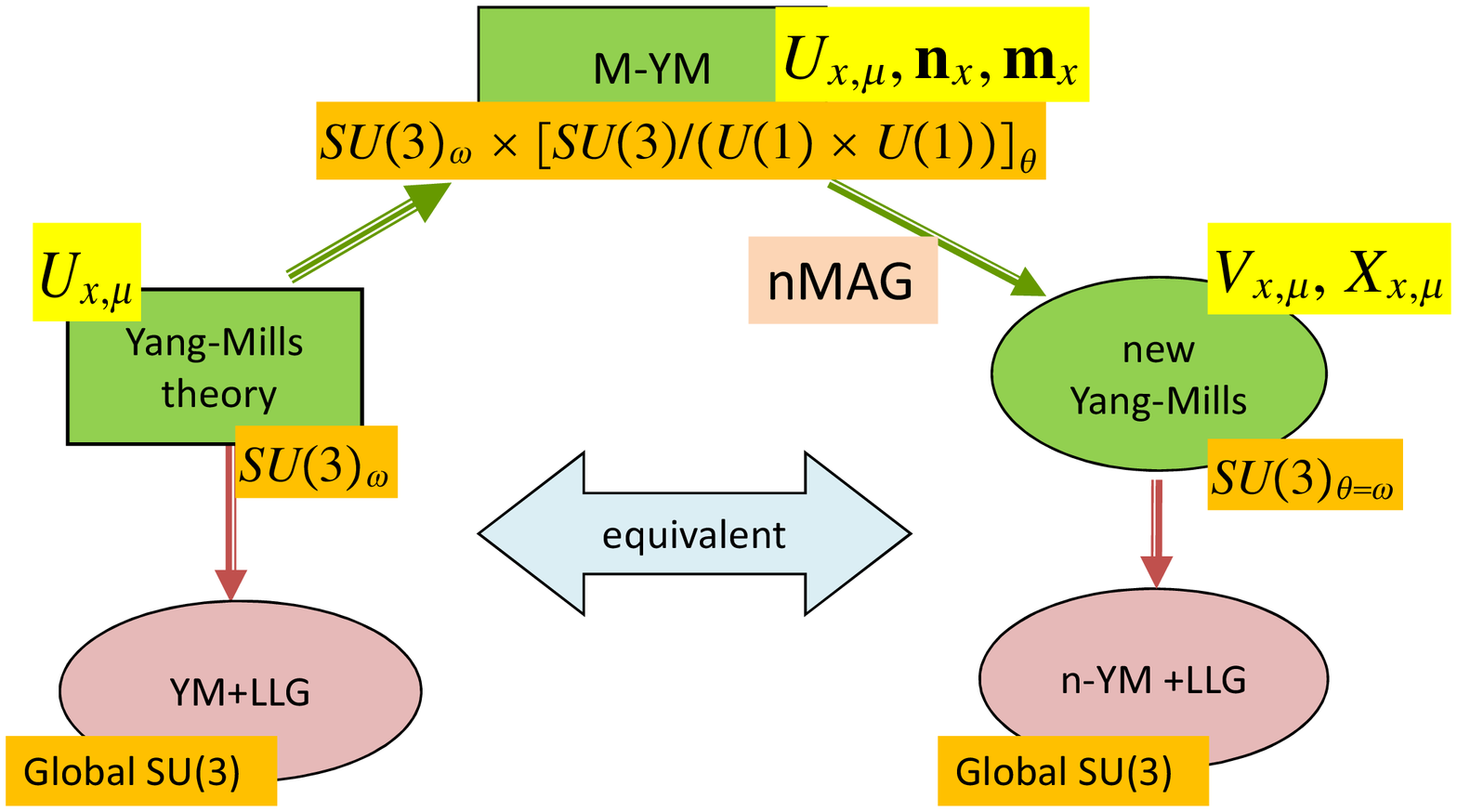}\includegraphics[
height=2.0 in, width=3.2 in
]{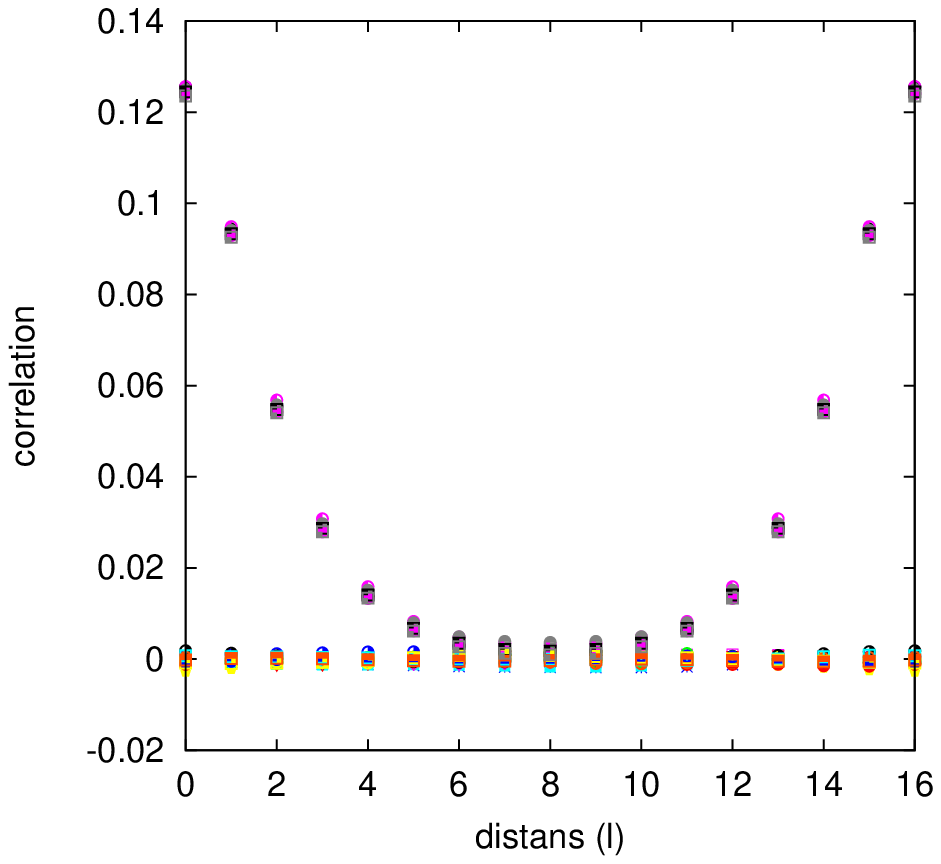}
\end{center}
\caption{(Left panel) The gauge symmetry of the new formulation. (Right panel)
The correlation functions for the color vector field $\mathbf{n}_{x}:$
$<n_{x,}^{A}n_{y}^{B}>$, where ($AB$) elemens ($A<B$) are plotted.}%
\label{fig:fig1}%
\end{figure}

We construct an $SU(3)$ lattice formulation by extending the $SU(2)$ case
\cite{ref:NLCVsu2,ref:NLCVsu2-2}. Two color vector fields, $\mathbf{n}%
_{\mathbf{x}}$ and $\mathbf{m}_{x}$, are introduced. They play a crucial role
of maintaining the color symmetry of the original YM theory. A link variable
$U_{x,\mu}$ represents exponential of the line integral of a gauge potential
$\mathbf{A}_{\mu}$ along a link from $x$ to $x+\epsilon\hat{\mu}$:%
\begin{equation}
U_{x,\mu}=\mathcal{P}\exp\left(  -ig\int_{x}^{x+\epsilon\mu}dx^{\mu}%
\mathbf{A}_{\mu}(x)\right)  =\exp(-ig\epsilon\mathbb{A}_{x^{\prime},\mu}),
\label{eq:U-link}%
\end{equation}
where $\epsilon$ denotes a lattice spacing and $\mathcal{P}$ the path ordering
operator. In explicitly estimating the naive continuum limit we adopt the
midpoint ($x^{\prime}=x+\epsilon\hat{\mu}/2$) definition for the link
variable. Then we obtain an extended theory, M-YM ($U_{x,\mu},$ $\mathbf{n}%
_{x,},\mathbf{m}_{x}$), which has an enlarged gauge symmetry $SU(3)_{\omega
}\times\lbrack SU(3)/U(1)^{2}]_{\theta}$ (See Figure \ref{fig:fig1} ). Under
the gauge transformation $\Theta_{x}=\exp(i\mathbf{\theta}_{x}),$ $\Omega
_{x}=\exp(i\mathbf{\omega}_{x})\in SU(3)$, the color fields $\mathbf{n}%
_{\mathbf{x}}$, $\mathbf{m}_{x}$ and $U_{x,\mu}$\ transform as
\begin{align}
\mathbf{n}_{x}  &  \rightarrow{}^{\Theta}\mathbf{n}_{x}=\Theta_{x}%
\mathbf{n}_{x}\Theta_{x}^{\dag},\quad\mathbf{m}_{x}\rightarrow{}^{\Theta
}\mathbf{m}_{x}=\Theta_{x}\mathbf{m}_{x}\Theta_{x}^{\dag},\label{eq:GT-mn}\\
U_{x,\mu}  &  \rightarrow{}^{\Omega}U_{x,\mu}=\Omega_{x}U_{x,\mu}\Omega
_{x+\mu}^{\dag} \label{eq:GT-U}%
\end{align}
New compact lattice variables $V_{x,\mu}$ and $X_{x,\mu}$ which correspond to
the decomposed variables in continuum theory, $\mathbf{V}_{\mu}$ and
$\mathbf{X}_{\mu}$, should be represented by $\mathbf{n}_{\mathbf{x}}$,
$\mathbf{m}_{x}$ and $U_{x,\mu}$ in a similar way to the $SU(2)$ case. The
lattice version of the decomposition condition is given by
\begin{align}
D_{\mu}^{\epsilon}[V_{x,\mu}]\mathbf{n}_{x}  &  =0,\qquad D_{\mu}^{\epsilon
}[V_{x,\mu}]\mathbf{m}_{x}=0,\label{eq:NLCV-01}\\
\mathrm{Tr}(\mathbf{n}_{x}X_{x,\mu})  &  =0,\qquad\mathrm{Tr}(\mathbf{m}%
_{x}X_{x,\mu})=0, \label{eq:NLCV-02}%
\end{align}
where $D_{\mu}^{\epsilon}[V_{x,\mu}]\mathbf{\phi}_{x}$ is a lattice version of
the covariant derivative defined by
\begin{equation}
D_{\mu}^{\epsilon}[V_{x,\mu}]\mathbf{\phi}_{x}:=\frac{1}{\epsilon}\left(
V_{x,\mu}\phi_{x+\mu}-\phi_{x}V_{x,\mu}\right)  .
\end{equation}
Under the gauge transformation
\begin{align}
&  V_{x,\mu}\rightarrow{}^{\Theta}V_{x,\mu}=\Theta_{x}V_{x,\mu}\Theta_{x+\mu
}^{\dag},\text{ }\label{eq:GT-V}\\
&  X_{x,\mu}\rightarrow{}^{\Theta}X_{x,\mu}=\Theta_{x}X_{x,\mu}\Theta
_{x}^{\dag}, \label{eq:GT-X}%
\end{align}
the decomposition condition satisfies desired gauge transformations, $D_{\mu
}^{\epsilon}[V_{x,\mu}]\mathbf{\phi}_{x}\rightarrow$ $D_{\mu}^{\epsilon}%
[{}^{\Theta}V_{x,\mu}]{}^{\Theta}\mathbf{\phi}_{x}$ $=\Theta_{x}\left[
D_{\mu}^{\epsilon}[V_{x,\mu}]\mathbf{\phi}_{x}\right]  \Theta_{x+\mu}^{\dag}$,
$\mathrm{Tr}(\mathbf{n}_{x}X_{x,\mu})=\mathrm{Tr}({}^{\Theta}\mathbf{n}_{x}%
{}^{\Theta}X_{x,\mu})$ and $\mathrm{Tr}(\mathbf{m}_{x}X_{x,\mu})=\mathrm{Tr}%
({}^{\Theta}\mathbf{m}_{x}{}^{\Theta}X_{x,\mu})$. To define the equivalent
theory in terms of the new variables to the original YM, the extended symmetry
should be restricted to the same symmetry as the original YM, $SU(3)_{\omega
=\theta}$. For this purpose, we use a new MAG (nMAG) condition which is
obtained by minimizing the functional:%
\begin{align}
&  F_{\mathrm{nMAG}}[\Omega,\Theta;U_{x,\mu},\mathbf{n}_{x},\mathbf{m}%
_{x}]=\sum_{x,\mu}\left\{  \left\Vert D_{\mu}^{\epsilon}[^{\Omega}U_{x,\mu
}]^{\Theta}\mathbf{n}_{x}\right\Vert ^{2}+\left\Vert D_{\mu}^{\epsilon
}[^{\Omega}U_{x,\mu}]^{\Theta}\mathbf{m}_{x}\right\Vert ^{2}\right\}
\label{eq:nMAG}\\
&  =\sum_{x,\mu}Tr({}^{\Omega}U_{x,\mu}{}^{\Theta}\mathbf{n}_{x+\mu}{}%
^{\Omega}U_{x,\mu}^{-1}{}^{\Theta}\mathbf{n}_{x})+\sum_{x,\mu}Tr({}^{\Omega
}U_{x,\mu}{}^{\Theta}\mathbf{m}_{x+\mu}{}^{\Omega}U_{x,\mu}^{-1}{}^{\Theta
}\mathbf{m}_{x})+c.c.\nonumber
\end{align}

Since (\ref{eq:NLCV-01}) and (\ref{eq:NLCV-02}) are coupled matrix equations,
it is difficult to obtain the general solution. Therefore, we consider a
formula which reproduce the NLCV in the continuum theory. We adopt the
midpoint definition for $V_{x,\mu}$ and the site definition for $X_{x,\mu}$:
\begin{align}
V_{x,\mu}  &  =\exp(-ig\epsilon\mathbb{V}_{x^{\prime},\mu})=\mathcal{P}%
\exp\left(  -ig\int_{x}^{x+\mu}dx^{\mu}\mathbf{V}_{\mu}(x)\right)
,\label{eq:link-V}\\
X_{x,\mu}  &  =\exp(-ig\epsilon\mathbb{X}_{x,\mu}), \label{eq:link-X}%
\end{align}
where $V_{x,\mu}$ could be the link variable represented by exponential of the
line integral of $\mathbf{V}_{\mu}$ like $U_{x,\mu}$. \ In the naive continuum
limit $D_{\mu}^{\epsilon}[V_{x,\mu}]\mathbf{\phi}_{x}$ agrees with the
continuum version up to $O(\epsilon^{2}):$
\begin{equation}
D_{\mu}^{\epsilon}[V_{x,\mu}]\mathbf{\phi}_{x}=\partial_{\mu}\mathbf{\phi
}_{x^{\prime}}-ig\left[  \mathbb{V}_{x^{\prime},\mu},\mathbf{\phi}_{x^{\prime
}}\right]  +\frac{ig\epsilon}{2}\left\{  \partial_{\mu}\mathbf{\phi
}_{x^{\prime}}-ig\left[  \mathbb{V}_{x^{\prime},\mu},\mathbf{\phi}_{x^{\prime
}}\right]  ,\mathbb{V}_{x^{\prime},\mu}\right\}  +O(\epsilon^{2}).
\end{equation}
We take an ansatz;
\begin{align}
\tilde{V}_{x,\mu}  &  =\alpha U_{x,\mu}+\beta_{1}\mathbf{n}_{x}U_{x,\mu}%
+\beta_{2}\mathbf{m}_{x}U_{x,\mu}+\beta_{3}U_{x,\mu}\mathbf{n}_{x+\mu}%
+\beta_{4}U_{x,\mu}\mathbf{m}_{x+\mu}\nonumber\\
&  +\gamma_{1}\mathbf{n}_{x}U_{x,\mu}\mathbf{n}_{x+\mu}+\gamma_{2}%
\mathbf{m}_{x}U_{x,\mu}\mathbf{m}_{x+\mu}+\gamma_{3}\mathbf{n}_{x}U_{x,\mu
}\mathbf{m}_{x+\mu}+\gamma_{4}\mathbf{n}_{x}U_{x,\mu}\mathbf{m}_{x+\mu
},\label{Anz:V}\\
V_{x,\mu}  &  =P_{x,\mu}^{-1}\tilde{V}_{x,\mu},\qquad P_{x,\mu}:=\sqrt
{\tilde{V}_{x,\mu}\tilde{V}_{x,\mu}^{\dag}},\label{def:V}\\
X_{x,\mu}  &  :=U_{x,\mu}V_{x,\mu}^{\dag}, \label{def:X}%
\end{align}
where the variable $\tilde{V}_{x,\mu}$ is represented by polynomials of
$U_{x,\mu},$\ $\mathbf{m}_{x}$ and $\mathbf{n}_{x}$ and should satisfy the
property of the gauge transformation (\ref{eq:GT-mn}), (\ref{eq:GT-U}) and
(\ref{eq:GT-V}). Here (\ref{def:V}) represents the polar decomposition for
obtaining the unitary matrix $V_{x,\mu}$ from $\tilde{V}_{x,\mu}$, where
$P_{x,\mu}$ is a Hermitian matrix, $P_{x,\mu}^{\dag}=P_{x,\mu}.$ By
substituting (\ref{Anz:V}) into (\ref{eq:NLCV-01}), the coefficients of
$\tilde{V}_{x,\mu}$ are determined as
\begin{equation}
\tilde{V}_{x,\mu}=\alpha U_{x,\mu}+\gamma\mathbf{n}_{x}U_{x,\mu}%
\mathbf{n}_{x+\mu}+\gamma\mathbf{m}_{x}U_{x,\mu}\mathbf{m}_{x+\mu}.
\label{eq:V}%
\end{equation}
Note that (\ref{Anz:V}) is used instead of $V_{x,\mu}$ to determine the
coefficients as a necessary condition, since the relation $D_{\mu}^{\epsilon
}[V_{x,\mu}]\mathbf{n}_{x}=0$ is obtained if $D_{\mu}^{\epsilon}[\tilde
{V}_{x,\mu}]\mathbf{n}_{x}=0$ is satisfied. Then, by using (\ref{def:X}) and
(\ref{eq:NLCV-02}), the coefficients in (\ref{eq:V}) are determined as
$\gamma=6\alpha$ up to $O(\epsilon)$ for $\mathbb{X}_{x^{\prime},\mu}$, and
the overall factor $\alpha$ can be set to $\alpha=1.$ In the continuum limit,
$\mathbb{V}_{x^{\prime},\mu}$ is given by%
\begin{equation}
\mathbb{V}_{x^{\prime},\mu}=\mathrm{Tr}(\mathbb{A}_{x^{\prime},\mu}%
\mathbf{n}_{x^{\prime}})\mathbf{n}_{x^{\prime}}+\mathrm{Tr}(\mathbb{A}%
_{x^{\prime},\mu}\mathbf{m}_{x^{\prime}})\mathbf{m}_{x^{\prime}}+1/g\left[
\partial_{\mu}\mathbf{n}_{x^{\prime}},\mathbf{n}_{x^{\prime}}\right]
+1/g\left[  \partial_{\mu}\mathbf{m}_{x^{\prime}},\mathbf{m}_{x^{\prime}%
}\right]  ,
\end{equation}
which agrees with $\mathbf{V}_{\mu}(x^{\prime})$ in the continuum theory.
$\mathbb{X}_{\mu}$ can be defined in two ways; one is a definition on the
midpoint, $\mathbb{X}_{x^{\prime},\mu}$ $=\mathbb{A}_{x^{\prime},\mu
}-\mathbb{V}_{x^{\prime},\mu}$, and the other $\mathbb{X}_{x,\mu}$ on a
lattice site from $X_{x,\mu}$ in terms of $\mathbb{A}_{x^{\prime},\mu}$and
$\mathbb{V}_{x^{\prime},\mu}$:%
\[
X_{x,\mu}=\exp(-ig\epsilon\mathbb{X}_{x,\mu})=U_{x,\mu}V_{x,\mu}^{\dag}%
=\exp\left(  -ig\epsilon\left(  \mathbb{A}_{x^{\prime},\mu}-\mathbb{V}%
_{x^{\prime},\mu}\right)  -\frac{g^{2}\epsilon^{2}}{2}\left[  \mathbb{A}%
_{x^{\prime},\mu},\mathbb{V}_{x^{\prime},\mu}\right]  +o(\epsilon^{3})\right)
.
\]

\section{Determining the new variables from lattice data}

In this section, we consider a procedure for obtaining $V_{x,\mu}$ and
$X_{x,\mu}$ from numerical simulations. To calculate $P_{x,\mu}$, in general,
we need to diagonalize $P_{x,\mu}^{2}=\tilde{V}_{x,\mu}\tilde{V}_{x,\mu}%
^{\dag}$ by solving an eigenvalue problem. Here we discuss the way to obtain
$V_{x,\mu}$ using the nMAG\ condition.

Suppose a gauge transformation $\Theta_{x}$ $\in SU(3)$ diagonalizes \ the
color vector fields $\mathbf{n}_{x}$ and $\mathbf{m}_{x}$ such that
$\Theta_{x}\mathbf{n}_{x}\Theta_{x}^{\dag}=\lambda^{3}$ and $\Theta
_{x}\mathbf{m}_{x}\Theta_{x}^{\dag}=\lambda^{8}.$ This is always possible
since $[\mathbf{n}_{x},\mathbf{m}_{x}]=0$ is satisfied. Using this gauge
transformation $U_{x,\mu}$ is transformed as $^{\Theta}U_{x,\mu}=\Theta
_{x}U_{x,\mu}\Theta_{x+\mu}^{\dag}$, and $V_{x,\mu}$\ and $P_{x,\mu}$ can also
be diagonalized at the same time, since we obtain the relations $[P_{x,\mu
},\mathbf{m}_{x}]=0$, $[P_{x,\mu},\mathbf{n}_{x}]=0$\ using the relations
$V_{x,\mu}\mathbf{m}_{x}=\mathbf{m}_{x}V_{x,\mu}$ and $V_{x,\mu}\mathbf{n}%
_{x}=\mathbf{n}_{x}V_{x,\mu}$ from (\ref{eq:NLCV-01}). Indeed, we can rewrite
(\ref{eq:V}) to a diagonalized form%
\begin{equation}
{}^{\Theta}\tilde{V}_{x,\mu}=\Theta_{x}\tilde{V}_{x,\mu}\Theta_{x+\mu
}^{\dagger}=\left[  ^{\Theta}U_{x,\mu}+6\lambda^{3}{}^{\Theta}U_{x,\mu}%
\lambda^{3}+\lambda^{8}{}^{\Theta}U_{x,\mu}\lambda^{8}\right]  =3\left[
\mathrm{diag}({}^{\Theta}u_{x,\mu}^{11},{}^{\Theta}u_{x,\mu}^{22},{}^{\Theta
}u_{x,\mu}^{33})\right]  , \label{eq:Vtilde}%
\end{equation}
where $^{\Theta}u_{x,\mu}^{AB}$ denotes an $(AB)$ element of $^{\Theta
}U_{x,\mu}$. Then, we have the diagonalized form of $P_{x,\mu}$, and
$V_{x,\mu}$ is given by%
\begin{equation}
{}^{\Theta}V_{x,\mu}=\Theta_{x}\left[  P_{x,\mu}^{-1}\tilde{V}_{x,\mu}\right]
\Theta_{x+\mu}^{\dagger}=\mathrm{diag}\left(  \frac{{}^{\Theta}u_{x,\mu}^{11}%
}{\left\vert {}^{\Theta}u_{x,\mu}^{11}\right\vert },\frac{{}^{\Theta}u_{x,\mu
}^{22}}{\left\vert {}^{\Theta}u_{x,\mu}^{22}\right\vert },\frac{{}^{\Theta
}u_{x,\mu}^{33}}{\left\vert {}^{\Theta}u_{x,\mu}^{33}\right\vert }\right)  .
\label{eq:Vdiag}%
\end{equation}

Therefore, the decomposition of $U_{x,\mu}$ is reduced to the problem of
finding out such as $\Theta=\bar{\Theta}_{x}.$ The solution $\bar{\Theta}_{x}$
can be obtained by the nMAG\ condition. Since $F_{\mathrm{nMAG}}$ is invariant
under the local gauge transformation of $SU(3)_{\omega=\theta},$ the solution
of the nMAG\ condition selects a gauge orbit in the extended gauge symmetry.
Therefore we can always choose such a gauge transformation that diagonalizes
color fields, ${}^{\Theta}\mathbf{n}_{x}=\lambda_{3}$ and ${}^{\Theta
}\mathbf{m}_{x\text{ }}=\lambda_{8}$. Using this $\bar{\Theta}_{x}^{\dag},$
the nMAG\ condition is rewritten to the same expression as the conventional
MAG condition($G=\bar{\Theta}_{x}\Omega$):%
\begin{align}
F_{\mathrm{nMAG}}  &  =\sum_{x,\mu}Tr({}^{G}U_{x,\mu}{}\lambda^{3}{}%
^{G}U_{x,\mu}^{-1}{}\lambda^{3})+\sum_{x,\mu}Tr({}^{G}U_{x,\mu}{}\lambda^{8}%
{}^{G}U_{x,\mu}^{-1}{}\lambda^{8})+c.c.\nonumber\\
&  =\sum_{x,\mu}\left(  \left\vert {}^{G}u_{x,\mu}^{11}\right\vert
^{2}+\left\vert {}^{G}u_{x,\mu}^{22}\right\vert ^{2}+\left\vert {}^{G}%
u_{x,\mu}^{33}\right\vert ^{2}\right)  . \label{eq:nMAG0}%
\end{align}
Note that this nMAG condition does not fix the gauge of the original YM,
$^{\Omega}U_{x,\mu},$ but selects a gauge orbit along the local gauge symmetry
$SU(3)_{\omega=\theta}$. The conventional MAG condition corresponds to a
special gauge choice of ${}^{\bar{\Theta}}\mathbf{n}_{x}=\lambda^{3}$ and
${}^{\bar{\Theta}}\mathbf{m}_{x\text{ }}=\lambda^{8}$ on the gauge orbit. When
we choose an overall gauge condition of the original YM theory, for example,
the lattice Landau gauge, $^{\Omega}U_{x,\mu}=\bar{U}_{x,\mu}$, the
configurations of $\mathbf{n}_{x}$ and $\mathbf{m}_{x\text{ }}$are determined
using the gauge transformation $\bar{\Theta}_{x}$: $\bar{U}_{x,\mu}%
=\bar{\Theta}_{x}{}^{G}U_{x,\mu}\bar{\Theta}_{x+\mu}^{\dag}$,
\begin{align}
\mathbf{n}_{x}  &  =\bar{\Theta}_{x}^{\dag}\lambda_{3}\bar{\Theta}_{x}%
,\qquad\mathbf{m}_{x}=\bar{\Theta}_{x}^{\dag}\lambda_{8}\bar{\Theta}_{x},\\
V_{x,\mu}  &  =\bar{\Theta}_{x}^{\dag}\mathrm{diag}\left(  \frac{{}%
^{G}u_{x,\mu}^{11}}{\left\vert {}^{G}u_{x,\mu}^{11}\right\vert },\frac{{}%
^{G}u_{x,\mu}^{22}}{\left\vert {}^{G}u_{x,\mu}^{22}\right\vert },\frac{{}%
^{G}u_{x,\mu}^{33}}{\left\vert {}^{G}u_{x,\mu}^{33}\right\vert }\right)
\bar{\Theta}_{x+\mu}.
\end{align}

\section{Lattice data}

Numerical simulations are done using the standard Wilson action of SU(3) YM.
The configurations are generated on$\ $a $16^{4}$ lattice at $\beta=5.7$ using
the Cabibbo-Marinari heatbath algorithm\cite{ref:CabbiboMarinari}. After 5000
thermalizing sweeps with the cold start, 120 configurations are stored every
100 sweeps. We choose the lattice Landau gauge (LLG) for the overall gauge
fixing of the original YM theory. In gauge fixing procedure, we use the
over-relaxation algorithm to update link variables by using the gauge
transformation of SU(2) sub-groups in\ the SU(3) gauge transformation. In
order to avoid the lattice Gribov copy problem in the both LLG\ and nMAG
condition, we try to find out the configuration which absolutely minimizes the
gauge fixing functional. In the process of minimizing the gauge fixing
functional for $U_{x,\mu}$, we have prepared 16 replicas generated by random
gauge transformations form $U_{x,\mu}$, and among them we have selected the
configuration which have attained the least value of the functional.

First, we check color symmetry of our new formulation, which is the global
$SU(3)$ symmetry to be preserved in LLG. Under the global gauge
transformation, the gauge fixing functional of LLG, $F_{LLG}[g]=%
{\textstyle\sum\nolimits_{x,\mu}}
\mathrm{Tr}({}^{g}U_{x,\mu})$, is invariant, while the color fields
$\mathbf{n}_{x},\mathbf{m}_{x}$ change their directions. Therefore, we measure
the space-time average of the color vector fields $n_{x}^{A}=\mathrm{Tr(}%
\mathbb{\lambda}^{A}\mathbf{n}_{x}\mathrm{)}$ and $m_{x}^{A}=\mathrm{Tr(}%
\mathbb{\lambda}^{A}\mathbf{m}_{x}\mathrm{)}$ and the correlation functions.
The right panel of Figure \ref{fig:fig1} shows, for examples, the correlation
functions for $\mathbf{n}_{x}$. The lattice data show that the color symmetry
is preserved: $\left\langle n^{A}\right\rangle =0$, $\left\langle
m^{A}\right\rangle =0$ and
\[
\left\langle n_{x}^{A}n_{y}^{B}\right\rangle =\delta^{AB}D_{NN}(l),\quad
\left\langle m_{x}^{A}m_{y}^{B}\right\rangle =\delta^{AB}D_{MM}(l),\text{
}\left\langle n_{x}^{A}m_{y}^{B}\right\rangle =0,\text{ (}y=x+l\hat{\mu
}\text{, }\mu=4).
\]
Note that this preserving color symmetry is an advantage of our new formulation.

\begin{figure}[ptb]
\begin{center}
\includegraphics[
height=3in, width=1.8in, angle=-90
]{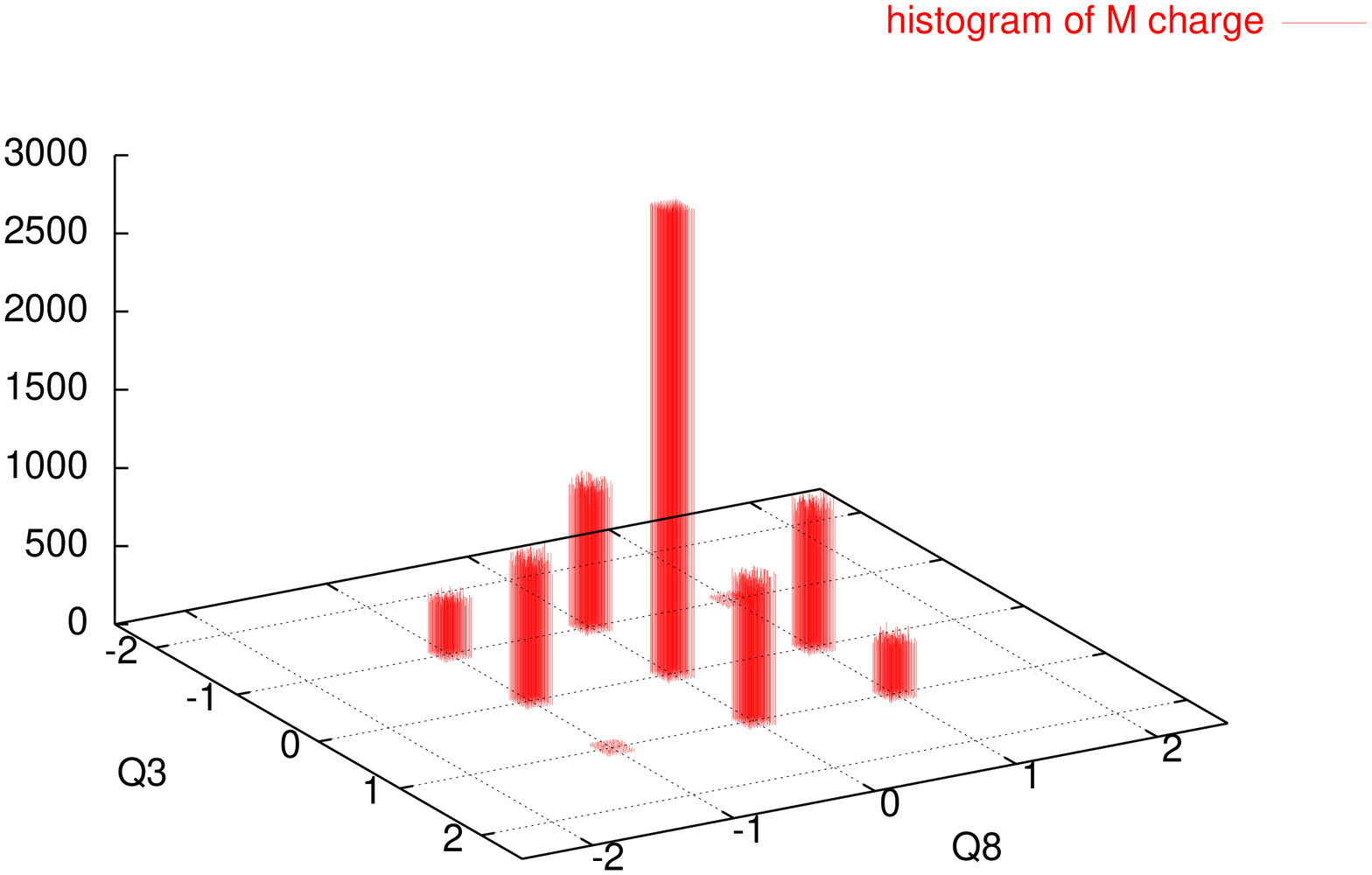}\includegraphics[
height=3in, width=1.8in, angle=-90
]{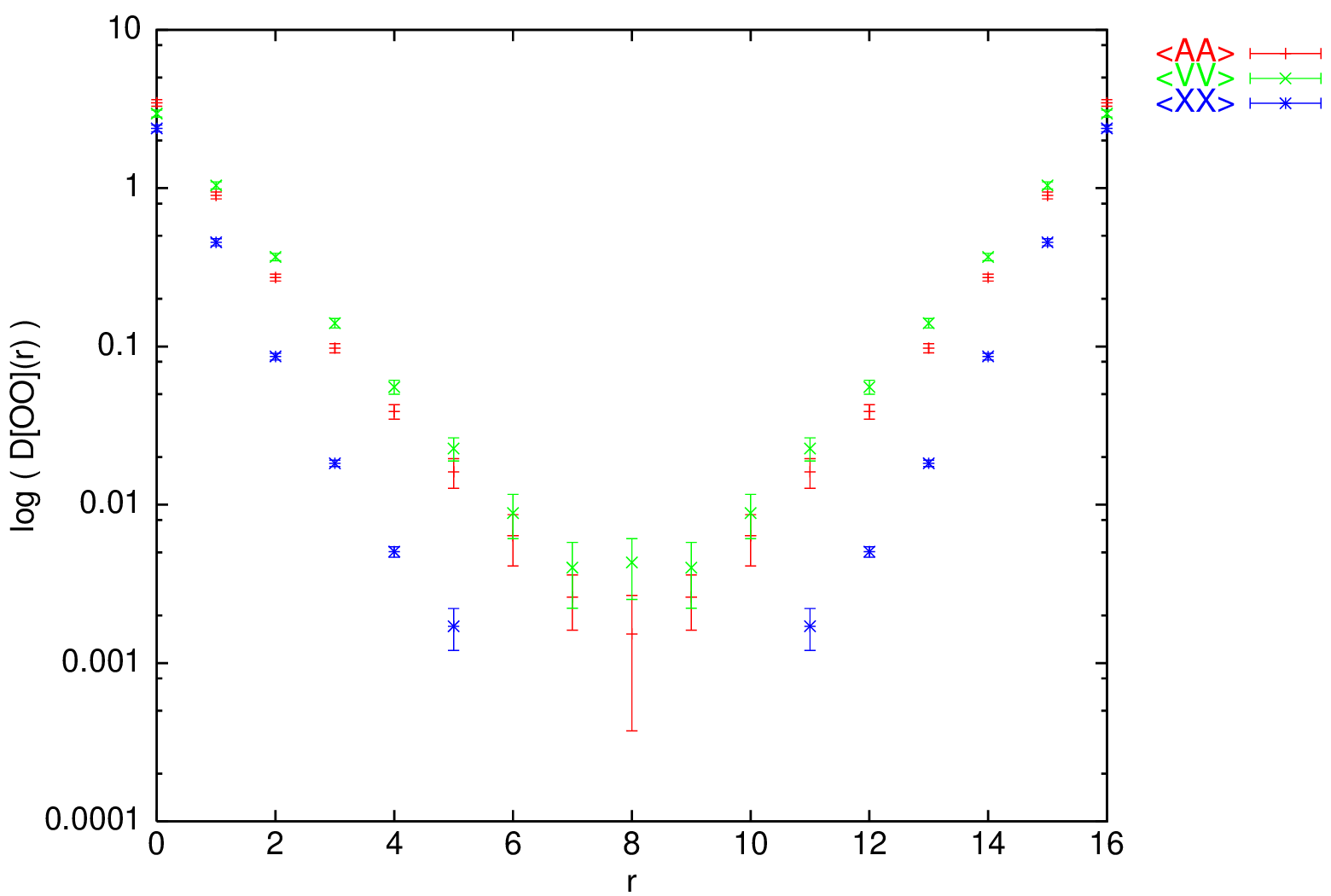}
\end{center}
\caption{(Left Panel) The histgram of monopole charges $Q^{(1)}$ and $Q^{(2)}%
$. The number of the monoples are plotted for 120 configurations. Each
distribution with integer valued monopole charges are represented on the
grids. This is a preliminary resut. (Right panel) The logarithm plot of
correlation functions of the gauge potential and the new variables: $D_{AA},$
$D_{VV}$ and $D_{XX}$.}%
\label{fig:fig20}%
\end{figure}

Next, we define a gauge invariant magnetic monopole using the "Abelian" part
$V_{x,\mu}$ in the similar way to the SU(2) case \cite{ref:NLCVsu2}. Two kinds
of the gauge invariant magnetic monopole currents ($\alpha=1,2$) are defined
by%
\begin{align}
k_{\mu}^{(a)}  &  :=\frac{1}{2}\epsilon_{\mu\nu\alpha\beta}\partial^{\nu
}\Theta_{\alpha\beta}^{(a)},\\
\Theta_{\alpha\beta}^{(1)}  &  :=\arg\mathrm{Tr}\left(  \left(  \frac{1}%
{3}I+\mathbf{n}_{x}+\frac{1}{\sqrt{3}}\mathbf{m}_{x}\right)  V_{x,\alpha
}V_{x+\alpha,\beta}V_{x+\beta,\alpha}^{-1}V_{x,\beta}^{-1}\right)  ,\\
\Theta_{\alpha\beta}^{(2)}  &  :=\arg\mathrm{Tr}\left(  \left(  \frac{1}%
{3}I-\frac{2}{\sqrt{3}}\mathbf{m}_{x}\right)  V_{x,\alpha}V_{x+\alpha,\beta
}V_{x+\beta,\mu}^{-1}V_{x,\beta}^{-1}\right)  .
\end{align}
The gauge invariance of $\Theta_{\alpha\beta}^{(a)\text{ }}$ is clear by
definition. Note that $\Theta_{\alpha\beta}^{(a)}$ is the $a$-th element of
the diagonalized expression of $V_{x,\mu}V_{x+\mu,\nu}V_{x+\nu,\mu}%
^{-1}V_{x,\nu}^{-1},$ i.e., $\mathrm{diag}($ $\exp(ig^{2}\epsilon^{2}%
\Theta_{\mu\nu}^{1})$, $\exp(ig^{2}\epsilon^{2}\Theta_{\mu\nu}^{2})$,
$\exp(ig^{2}\epsilon^{2}\Theta_{\mu\nu}^{3}))$. The left panel of Figure
\ref{fig:fig20} shows the histogram of the magnetic monopole charges,
indicating that integer valued magnetic monopoles are obtained.

Finally, we investigate the propagators of the new variables. The correlation
functions (propagators) of the new variables and the gauge potential of YM are
defined by
\begin{equation}
D_{OO}(x-y):=\left\langle \mathbb{O}_{\mu}^{A}(x)\mathbb{O}_{\mu}%
^{A}(y)\right\rangle \,\text{\ for }O_{\mu}^{A}(x^{\prime})=\mathbb{A}%
_{x^{\prime},\mu}^{A},\mathbb{V}_{x^{\prime},\mu}^{A}\text{, }\mathbb{X}%
_{x^{\prime},\mu}^{A}, \label{eq:corrF}%
\end{equation}
where an operator $O_{\mu}^{A}(x)$ is defined as a linear type, e.g.,
$\mathbb{A}_{x^{\prime},\mu}=\left(  U_{x,\mu}-U_{x,\mu}^{\dag}\right)
/\left(  2\varepsilon g\right)  .$ The right panel of Figure \ref{fig:fig20}
shows preliminary measurements of correlation functions of $D_{AA},$ $D_{VV}$
and $D_{XX}$. The correlation $D_{VV}$ corresponding to the "Abelian" part
dumps slowly and has almost the same dumping rate as $D_{AA}$, while the
$D_{XX}$ corresponding to the "off-diagonal" part dumps quickly$.$ This
suggests that the "Abelian" part of the gluon propagator is dominated in the
infrared region, the and mass generation of the "off-diagonal" gluon.

\section{Summary and discussion}

We have proposed a new compact lattice formulations of SU(3) YM theory as an
extension of the SU(2) case \cite{ref:NLCVsu20}\cite{ref:NLCVsu2}%
\cite{ref:NLCVsu2-2}. This formulation has enabled us to define a gauge
invariant magnetic monopole in the compact formulation which guarantees that
the magnetic charge is integer-valued and obeys the Dirac quantization
condition. We have shown that the new variables can be obtained in any gauge
of YM theory. It is crucial to introduce the color fields $\mathbf{n}(x)$ and
$\mathbf{m(x)}$ to maintain the color symmetry of the original YM theory. We
have performed the numerical simulations and measurements on a lattice. We
have shown that color symmetry is preserved for the new variables, and the
integer-valued gauge invariant magnetic monopoles are obtained. Though these
results are preliminary, the lattice data suggest infrared "Abelian" dominance
and mass generation of the "off-diagonal" gluon by investigating the
propagators of the new variables. The mass generation of gluon can be
investigated in the same way in the $SU(2)$ case, and these are under investigation.

Through these numerical simulations, we have shown that our new formulation
enables the gauge independent investigation of the confinement mechanism to
overcome the problems in the conventional studies based on a special gauge
such as the MA gauge. Further studies such as "Abelian" dominance, monopole
dominance and mass generation of gluons in LLG\ and also in the other gauges,
are important to establish the dual superconductivity picture.%

\section*{Acknowledgement}%

This work is supported by the Large Scale Simulation Program No.07-15 (FY2007)
of High Energy Accelerator Research Organization(KEK), and by Grant-in-Aid for
Scientific Research (C)18540251 from JSPS.

%



\begin{thebibliography}{9}                                                                                                %


\bibitem {ref:DsuperCond}Y. Nambu, PRD \textbf{10}, 4262 (1974); G. 't Hooft,
in: High Energy Physics, edited by A. Zichichi (Editorice Compositori,
Bologna, 1975); S. Mandelstam, Phys. Report \textbf{23}, 245 (1976); A.M.
Polyakov, Nucl. Phys. B \textbf{120}, 429 (1977).

\bibitem {ref:tHooft81}G.'t Hooft, Nucl.Phys. B\textbf{190} [FS3], 455 (1981).

\bibitem {ref:EzawaIwasaki}Z.F. Ezawa and A. Iwazaki, PRD\textbf{25}, 2681 (1982).

\bibitem {ref:Kondo}K.-I. Kondo, T. Murakami, T. Shinohara,
Prog.Theor.Phys.\textbf{115,} 201-216 (2006); Eur. Phys. J.C \textbf{42}, 475
(2005); K.-I. Kondo, PRD \textbf{74, }125003 (2006)

\bibitem {ref:NLCVsu20}S. Kato, K.-I. Kondo, T. Murakami, A. Shibata, T.
Shinohara and S. Ito, PLB\textbf{ 632,} 326-332 (2006)

\bibitem {ref:NLCVsu2}S. Ito, S. Kato, K.-I. Kondo, A. Shibata, T. Shinohara,
PLB \textbf{645}, 67-74 (2007)

\bibitem {ref:NLCVsu2-2}A. Shibata, S. Kato, K.-I. Kondo, T. Murakami, T.
Shinohara, S. Ito, PLB \textbf{653}, 101-108 (2007)

\bibitem {ref:prep01}A. Shibata, S. Kato, K.-I. Kondo, T. Murakami, T.
Shinohara and S. Ito, in preparation.

\bibitem {ref:CabbiboMarinari}N.Cabibbo and E. Marinari , PLB \textbf{119},
387-390 (1982)
\end{thebibliography}
\end{document}